\documentclass[journal=nalefd,manuscript=article]{achemso}
\usepackage{chemformula} % Formula subscripts using \ch{}
\usepackage[T1]{fontenc} % Use modern font encodings
\usepackage{amsmath,bm}
\usepackage{txfonts}
\usepackage[utf8]{inputenc}
\usepackage{color}
\usepackage{hyperref}
\hypersetup{colorlinks=true,linkcolor=blue,filecolor=blue,urlcolor=blue,citecolor=blue}
\title[An \textsf{achemso} demo]
  {Spin-Selective Second-Order Topological Insulators Enabling Cornertronics in 2D Altermagnets\footnote{A footnote for the title}}

\author{Ning-Jing Yang}%$^{1,2}$
\affiliation{ Fujian Provincial Key Laboratory of Quantum Manipulation and New Energy Materials, College of Physics and Energy, Fujian Normal University, Fuzhou 350117, China}
\author{Zhigao Huang}%
\affiliation{ Fujian Provincial Key Laboratory of Quantum Manipulation and New Energy Materials, College of Physics and Energy, Fujian Normal University, Fuzhou 350117, China}
\author{Jian-Min Zhang}
\affiliation{ Fujian Provincial Key Laboratory of Quantum Manipulation and New Energy Materials, College of Physics and Energy, Fujian Normal University, Fuzhou 350117, China}
\email{jmzhang@fjnu.edu.cn}

\abbreviations{abb1, abb2}
\begin{document}

\begin{tocentry}
	\begin{center}
		\includegraphics[width=0.6\textwidth]{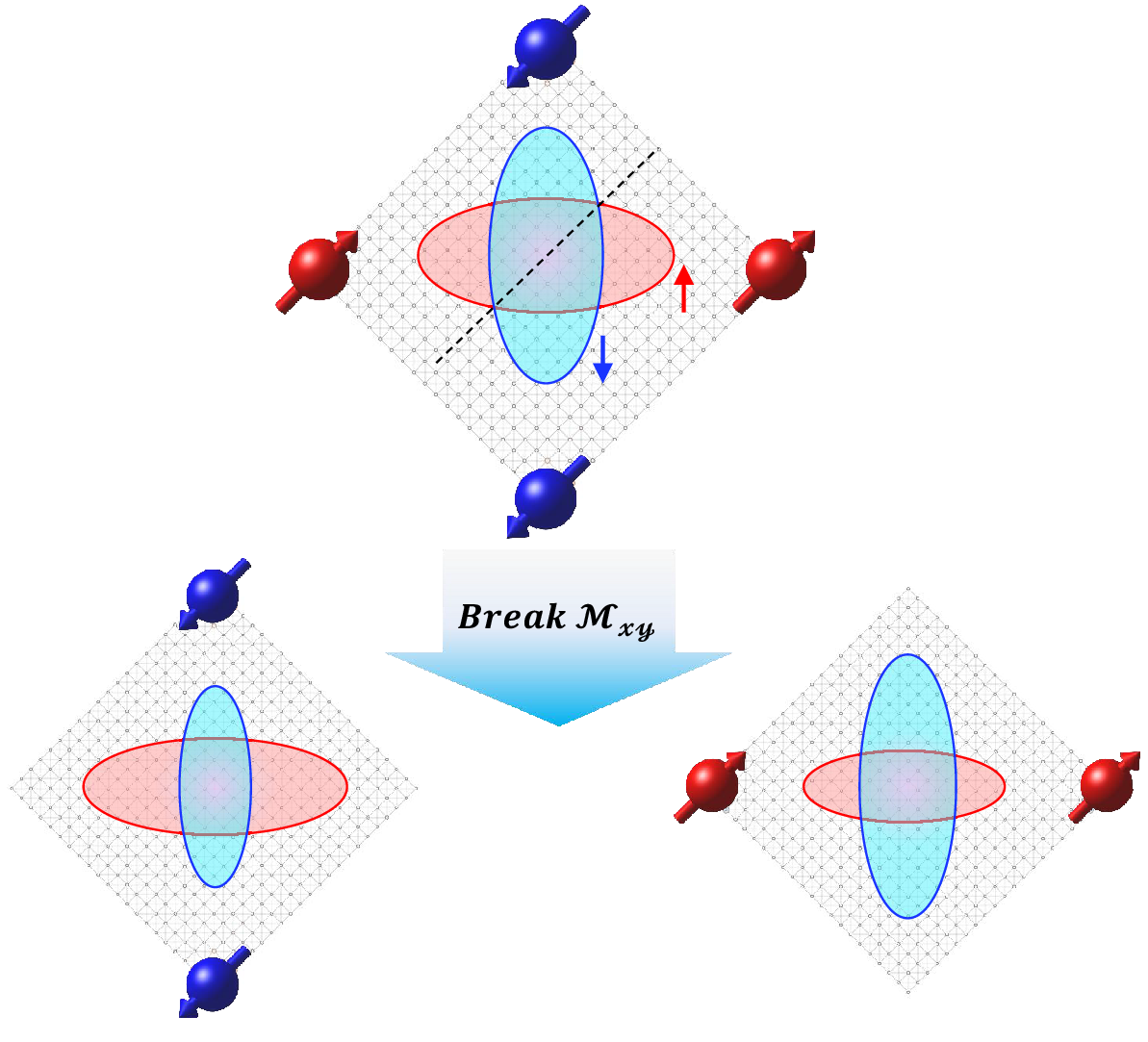}
	\end{center}
	
\end{tocentry}

\begin{abstract}
Recent progress in spintronics within the paradigm of altermagnets (AMs) opens new avenues for next-generation electronic device design. Here, we establish a spin-corner locking mechanism that generates second-order topological states in two-dimensional (2D) altermagnetic systems, through effective model analysis. Remarkably, the breaking of $M_{xy}$ symmetry under uniaxial strain creates spin-resolved corner modes, driving the system into a corner-polarized second-order topological insulator (CPSOTI). Beyond critical strain, a topological phase transition to quantum anomalous Hall insulator occurs with quantized conductance. Through first-principles calculations, we identify two experimentally viable candidates for 2D intrinsic AM CrO and Cr$_2$Se$_2$O - which host robust CPSOTI. Moreover, we construct the topological phase diagram of CrO and predict the existence of an altermagnetic Weyl semimetal phase. Our findings open technological avenues in altermagnetism and higher-order topology, while providing opportunities for coupling topological spintronics with cornertronics.

\end{abstract}

%\keywords{topological, American Chemical Society, \LaTeX}

%\keywords{High-order Topological,  \LaTeX}
%\section{Introduction}
\vspace{8mm}
Altermagnetism, an emerging third type of magnetic ordering, has recently attracted broad interest in spintronics and topological physics due to its intrinsic spin splitting \cite{PhysRevX.12.040501, PhysRevX.12.031042, song2025altermagnets, PhysRevLett.133.106701, PhysRevLett.132.263402, krempasky2024altermagnetic, PhysRevLett.130.216701}. Unlike conventional antiferromagnets, which exhibit spin-degenerate bands despite a zero net magnetization, altermagnets display spin-split band structures in momentum space. This spin splitting originates from spin-group symmetries rather than spin–orbit coupling, leading to an unconventional spin–momentum locking effect.
To date, numerous AMs have been theoretically predicted and experimentally realized, such as bulk materials RuO$_2$ and MnTe \cite{krempasky2024altermagnetic, PhysRevLett.130.216701}, as well as 2D compounds like V$_2$Se$_2$O and CaMnSi \cite{ma2021multifunctional, wang2024electric}. In 2D AMs, the emergence of non-hexagonal valleytronic features gives rise to spin-valley coupling, which can induce both an anomalous valley Hall effect and first-order topological phases \cite{jiang2025strain,PhysRevB.111.134429, li2024strain, li2024ferrovalley}.
Topological materials with nontrivial band structures have long been a central focus in condensed matter physics \cite{burkov2016topological, RevModPhys.93.025002, PhysRevLett.106.106802, tokura2019magnetic, PhysRevLett.122.206401}. In recent years, research has extended from conventional first-order topological phases to higher-order topological materials \cite{ hoti1, hoti2, hoti3, hoti4, ezawa2019, soti_ynj, PhysRevLett.130.116204}.
However, studies and in-depth discussions on higher-order topology in altermagnetic systems remain scarce.

In two-dimensional systems, higher-order topology is typically manifested as second-order corner states. 
However, the corner-related degrees of freedom are often fixed, and controllable mechanisms for corner states remain rare \cite{PhysRevLett.130.116204, zhan2024design, li2022robust, PhysRevB.107.235406}. The concept of tunable corner modes was first introduced in phononic crystals, where valley degrees of freedom (d.o.f.) enable selective corner states \cite{PhysRevLett.126.156401}. In electronic systems, the notion of "cornertronics" has only recently emerged, marking the beginning of controllable corner d.o.f. in electron-based materials \cite{PhysRevLett.133.176602, gong2024hidden}.
In 2D systems, Han et al. introduced the concept of cornertronics, where corner polarization is mediated by the valley degree of freedom in nonmagnetic systems \cite{PhysRevLett.133.176602}. However, in magnetic systems, the spin degree of freedom can serve as an alternative handle to regulate cornertronics. In this context, 2D altermagnets provide a naturally compatible platform, enabling spin-mediated higher-order topological corner states. Such phenomena highlight the need to unravel the interplay between multiple d.o.f. in complex quantum systems. Therefore, it is essential to explore higher-order topological cornertronics in altermagnetic systems featuring entangled spin, momentum, and corner.

In this Letter, we propose a universal and robust strategy to realize spin-resolved second-order topology in 2D altermagnets. Through an effective model, we uncover the microscopic origin of this spin-selective second-order topological phase, which stems from the intrinsic symmetries of 2D AMs. First-principles calculations on realistic candidates, CrO and Cr$_2$Se$_2$O, validate the emergence of spin-polarized corner states protected by the C$_{2z}$ symmetry. By selectively breaking M$_{xy}$ mirror symmetry while preserving C$_{2z}$, uniaxial strain enables controllable spin splitting and energy-level isolation of corner states. Our findings highlight a previously unexplored spin–corner–momentum locking mechanism, marking a distinct topological signature in altermagnetic systems. This work not only expands the classification of magnetic higher-order topological phases but also paves the way toward spin-resolved cornertronics in future quantum devices.

%%%%%%%%%%%%%%%%%%%%%%      k*p model    %%%%%%%%%%%%%%%%
\begin{figure*}[t]
	\centering
	\includegraphics[width= 16.0cm]{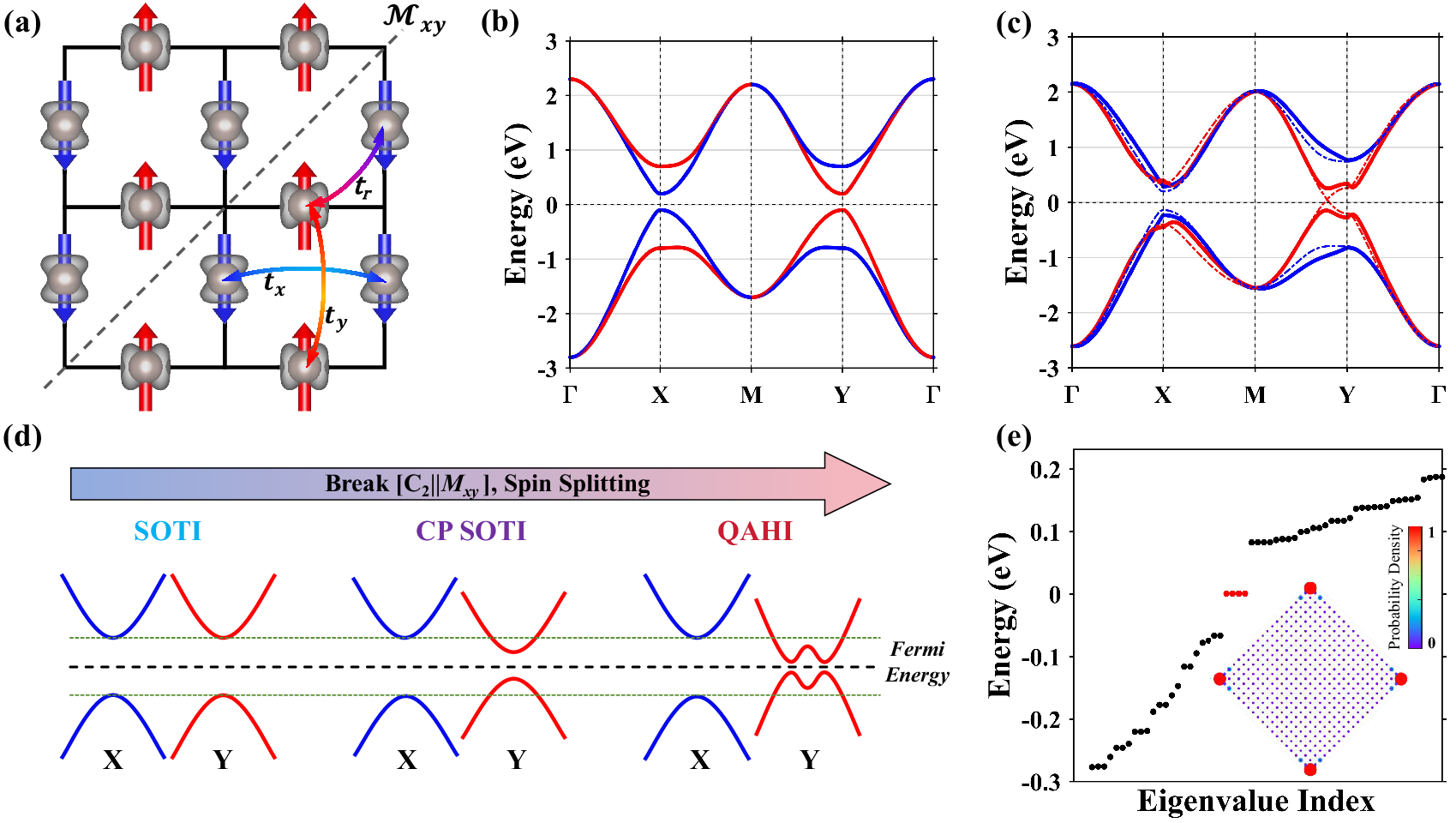}\caption{Effective model of second-order topological altermagnetic systems.
		(a) Illustration of the 2D AM SOTI model. 
		(b) AM band structure exhibiting spin splitting. 
		Here, $t_1^x=-1$ eV, $t_1^y=-1.5$ eV, $t_{2}^{x,y} = -0.8 t_1^{x,y}$, $t_r=1$.
		(c) Spin-polarized QAHI phase under broken \( M_{xy} \) symmetry. The dashed lines represent the results without considering SOC, while the solid lines correspond to the results with SOC included. Corresponds to $\delta = -0.1$.
		% \textcolor{red}{Corresponds to $\delta=−0.1$}.
		(d) Topological phase transition process under broken \( M_{xy} \) symmetry, wherein the AMs evolves from a SOTI phase to CPSOTI and QAHI phases. 
		(e) Energy spectrum on a quadrilateral quantum dot in the second-order topological phase, where a set of four corner states (colored dots) can be observed. The inset displays the real-space distribution of these corner states.
	}\label{pho:1}
\end{figure*}
%\vspace{-2pt}  % 减少图像下方的空间
%%%%%%%%%%%%%%%%%%%%%%%%%%%%%%%%%%%%%%%%%%%%%%%%%%%%%%%%%%

%%%%%%%%%%%%%%%%%%%     Preliminary discussion     %%%%%%%%%%%%%%%%%%%%%%%%%
%\section*{Approach $\&$ Model}
To elucidate the origin of second-order topology in 2D altermagnets, we first construct a minimal lattice model for a monolayer AM system. Without loss of generality, we assume that the AM system crystallizes in a square lattice, as shown in Fig.~\ref{pho:1} (a). The model incorporates two sublattices, and respects the spatial mirror symmetry \( M_{xy} \). Within the spin space group formalism, the system hosts a combined symmetry \([C_2 \parallel C_{4z}]\), which induces anisotropic spin splitting characteristic of altermagnetic systems, as shown in Fig.~\ref{pho:1} (b). Specifically, we develop a minimal tight-binding model based on the symmetry of spin space group (SSG) \( P^{\text{-}1}4/^1m^1m^{\text{-}1}m^{\infty m}1 \) to capture the second-order topological phase in this 2D square-lattice system. %It is important to note that this tight-binding model shares the same crystalline symmetries as those in 2D altermagnetic thin films. 
More importantly, as discussed below, the model successfully captures the essential topological features and phase transitions of 2D altermagnetic band structures. In this context, %under the spin–orbital basis \(\{|\phi_A^\uparrow\rangle, |\phi_A^\downarrow\rangle, |\phi_B^\uparrow\rangle, |\phi_B^\downarrow\rangle\}\), 
the symmetry-allowed Hamiltonian can be expressed as 
\begin{equation}
	\begin{split}	
		\begin{aligned}
			H= & \ t_{r} \sum_{i, a_{j}}\left(c_{1, i}^{\dagger} c_{2, i+a_{j}}+\text { h.c. }\right) 
			 +\sum_{\alpha, i}\left(t_{\alpha}^{x} c_{\alpha, i}^{\dagger} c_{\alpha, i+a_{x}}+t_{\alpha}^{y} c_{\alpha, i}^{\dagger} c_{\alpha, i+a_{y}}+\text { h.c. }\right) 
			 +\Delta \sum_{\alpha, i}(-1)^{\alpha} c_{\alpha, i}^{\dagger} c_{\alpha, i},
			%+M \sum_{\alpha, i}(-1)^{\alpha} c_{\alpha, i}^{\dagger} c_{\alpha, i} s_{z}
		\end{aligned}
	\end{split}
\end{equation}
where \( t_r \) represents the nearest-neighbor hopping, \( t_x \) and \( t_y \) denote the hopping amplitudes along the \(x\) and \(y\) directions, respectively. Here, $\alpha = 1, 2$ denote the sublattices located at \((0.5, 0)\) and \((0, 0.5)\), respectively. \(\Delta\) is the staggered potential. %$M$ is the net magnetization, and $s_z$ is the spin operator. 
Notably, to preserve the \([C_2 \parallel C_{4z}]\) symmetry, the hopping parameters must satisfy the condition \( t_x^\uparrow = t_y^\downarrow \) (and \( t_y^\uparrow = t_x^\downarrow \)), which is essential for inducing spin-splitting associated with the altermagnetic character, as shown in Fig.~\ref{pho:1} (b). In contrast, if \( t_x^\uparrow = t_x^\downarrow \) and \( t_y^\uparrow = t_y^\downarrow \), the system instead preserves the \([C_2 \parallel P]\) symmetry, leading to spin-degenerate bands characteristic of a trivial antiferromagnet~\cite{PhysRevLett.134.116703_ChengchengLiu}.
In realistic 2D materials, external perturbations such as strain and stress frequently break the original crystalline symmetry. 
Under uniaxial strain, the SSG of the 2D AM transforms from $P^{\text{-}1}4/^1m^1m^{\text{-}1}m^{\infty m}1$ to $P^1m^1m^{\text{-}1}m^{\infty m}1$, and the space group changes from \textit{P4/mmm} to \textit{Pmmm}. Consequently, the mirror symmetry ($M_{xy}$) is broken, as are the SSG symmetries [$C_2 \parallel M_{xy}$] and [$C_2 \parallel C_{4z}$], driving the system toward a ferrimagnetic configuration and resulting in spin polarization. However, the $C_{2z}$ symmetry is preserved, which is essential for the SOTI phase.
In such cases, the emergence of a quantum anomalous Hall insulator (QAHI) becomes possible~\cite{guo2023quantum}. To realize QAHI, spin–orbit coupling (SOC) is indispensable, and many 2D AMs exhibit a finite SOC~\cite{guo2023quantum,jiang2025strain,PhysRevB.111.134429}.
Therefore, to capture these effects, we extend the Hamiltonian to include SOC and symmetry-breaking perturbations, which are expressed as follows:
\begin{equation}
	\begin{array}{l}
		H_{s o c}=\sum_{i, a_{j}} C_{1, i}^{\dagger}(i \lambda \cdot \boldsymbol{\sigma}) C_{2, i+\alpha j}^{\dagger} \\
		H_{\delta}=\delta\left(\cos \boldsymbol{k}_{x}-\cos \boldsymbol{k}_{y}\right) \boldsymbol{S}_{z} \otimes \mathrm{I}.
	\end{array}
\end{equation}
Although altermagnetism is typically associated with SOC-free scenarios, in the presence of certain forms of SOC, the system can still preserve symmetries associated with specific spin space groups~\cite{PhysRevX.14.031037,PhysRevX.14.031038,PhysRevX.14.031039}. For instance, when a weak lattice distortion (\( \delta = 0.1 \)) is introduced, spin polarization leads to a spin-selective Dirac point along the high-symmetry M–Y line. 
The band gap closure induced by symmetry breaking transforms the system into a nodal semimetal. 
Upon introducing SOC, a gap opens at this crossing point, transforming the system from a semimetal into a QAHI, as illustrated in Fig. ~\ref{pho:1} (c).
SOC is the key factor that induces the QAHI phase, but it does not affect the second-order topology, as it preserves the C$_{2z}$ symmetry.

\begin{figure*}[t]
	\centering
	\includegraphics[width=12cm]{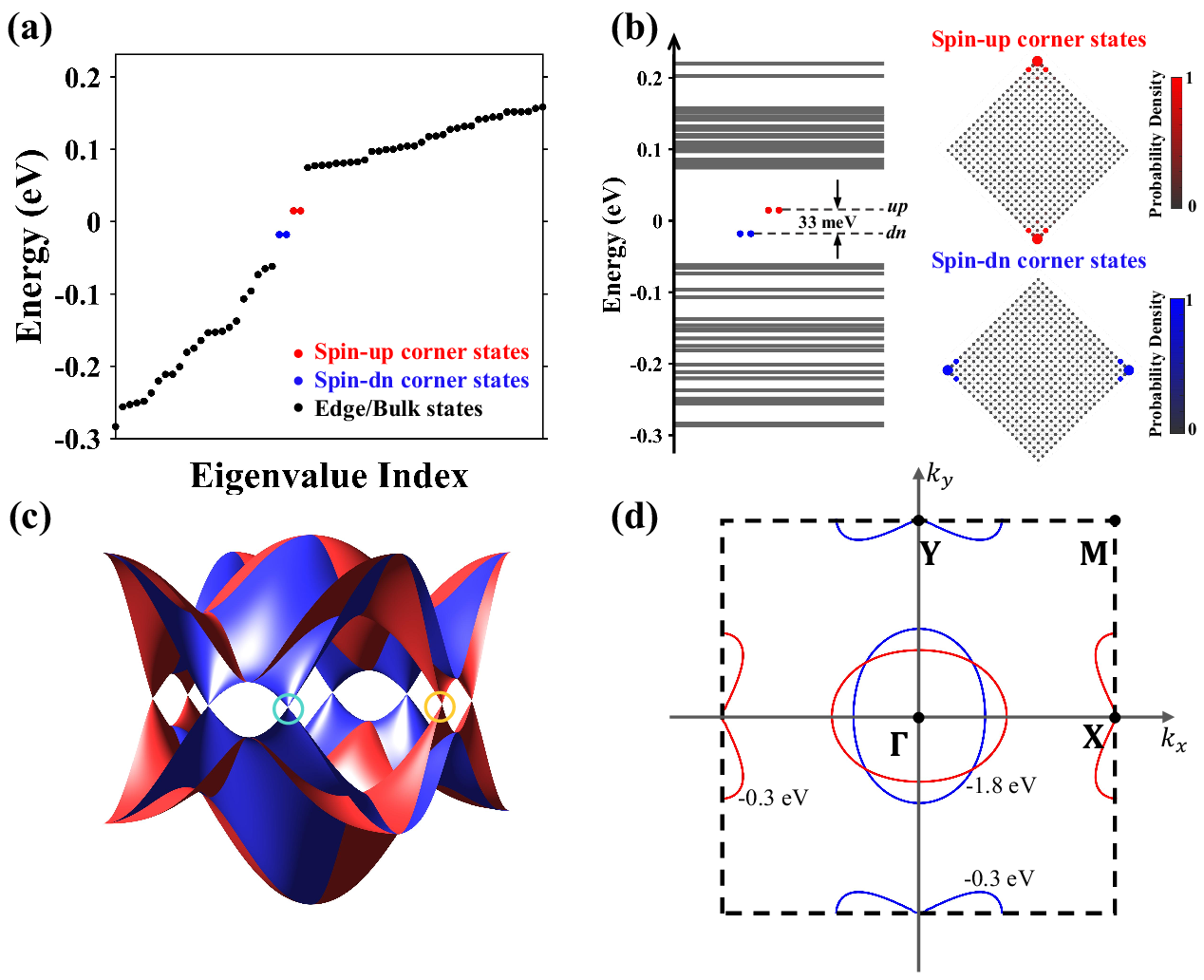}\caption{(a) Energy spectrum of a quantum dot with broken $M_{xy}$ symmetry in AMs, showing two sets of corner states (colored dots).
		(b) Corresponding energy level diagram of the quantum dot, where black horizontal lines denote bulk and edge states. Red and blue dots represent spin-resolved corner states, with their real-space projections shown on the right.
		(c) 3D spin-resolved band structure of the AM Weyl semimetal. Degenerate points at the Fermi level for different spin bands are marked with colored circles.
		(d) Fermi surfaces of the AM Weyl semimetal at different chemical potentials.}\label{pho:weyl}
\end{figure*}

Meanwhile, this effective model also captures the second-order topological characteristics. By constructing a quantum disk based on the square lattice, we identify four corner-localized states within the bulk energy gap, as shown in Fig.~\ref{pho:1} (e). 
When constructing square quantum dots with altermagnetism, it is also necessary to maintain the consistency of symmetry. In this system, the C$_{2z}$ symmetry with respect to the spin index must be preserved to ensure the emergence of corner states.
By calculating the electron filling in a charge-neutral nanodisk, we find that the fourfold-degenerate corner states are occupied by two electrons—one with spin-up and the other with spin-down. Each corner hosts a partially filled state, giving rise to a fractional corner charge of \( e/2 \). In fact, the corner states are spin-resolved, a feature that becomes more evident upon introducing symmetry-breaking perturbations. This leads to spin-polarized corner modes, as illustrated in Fig.~\ref{pho:weyl} (a). 
Similarly, the two spin channels form doubly degenerate corner states along the horizontal and vertical axes, respectively, as shown in Fig.~\ref{pho:weyl} (b). Each of these spin-polarized corner states also carries a \( e/2 \) corner charge. This configuration gives rise to a corner-polarized second-order topological insulator (CPSOTI), which emerges as a result of breaking the mirror symmetry \( M_{xy} \). Notably, as shown in Fig.~\ref{pho:weyl} (b), the energy splitting between the spin-resolved corner states can reach up to 33~meV, enabling experimental tuning of the corner charge via control of the chemical potential. In this regime, the corner degrees of freedom associated with the corner electrons become clearly distinguishable.
Therefore, during the symmetry-breaking process of \( M_{xy} \), a rare and intriguing CPSOTI phase arises between the SOTI and QAHI phases, as schematically illustrated in the phase diagram of Fig.~\ref{pho:1} (d).

In systems that preserve mirror symmetry and are free of spin-orbit coupling, staggered magnetic Weyl semimetals can also emerge in addition to gapped staggered magnetic insulators, as illustrated by the 3D band structure in Fig. \ref{pho:weyl} (c). By tuning the hopping amplitudes, spin bands degeneracy at the boundary of the first Brillouin zone. 
The corresponding Fermi surfaces at -0.3 eV and -1.8 eV are shown in Fig. \ref{pho:weyl} (d), highlighting the distinctive spin-splitting characteristic of AMs.

%\section*{Material realization}

\begin{figure*}[t]
	\centering
	\includegraphics[width=16cm]{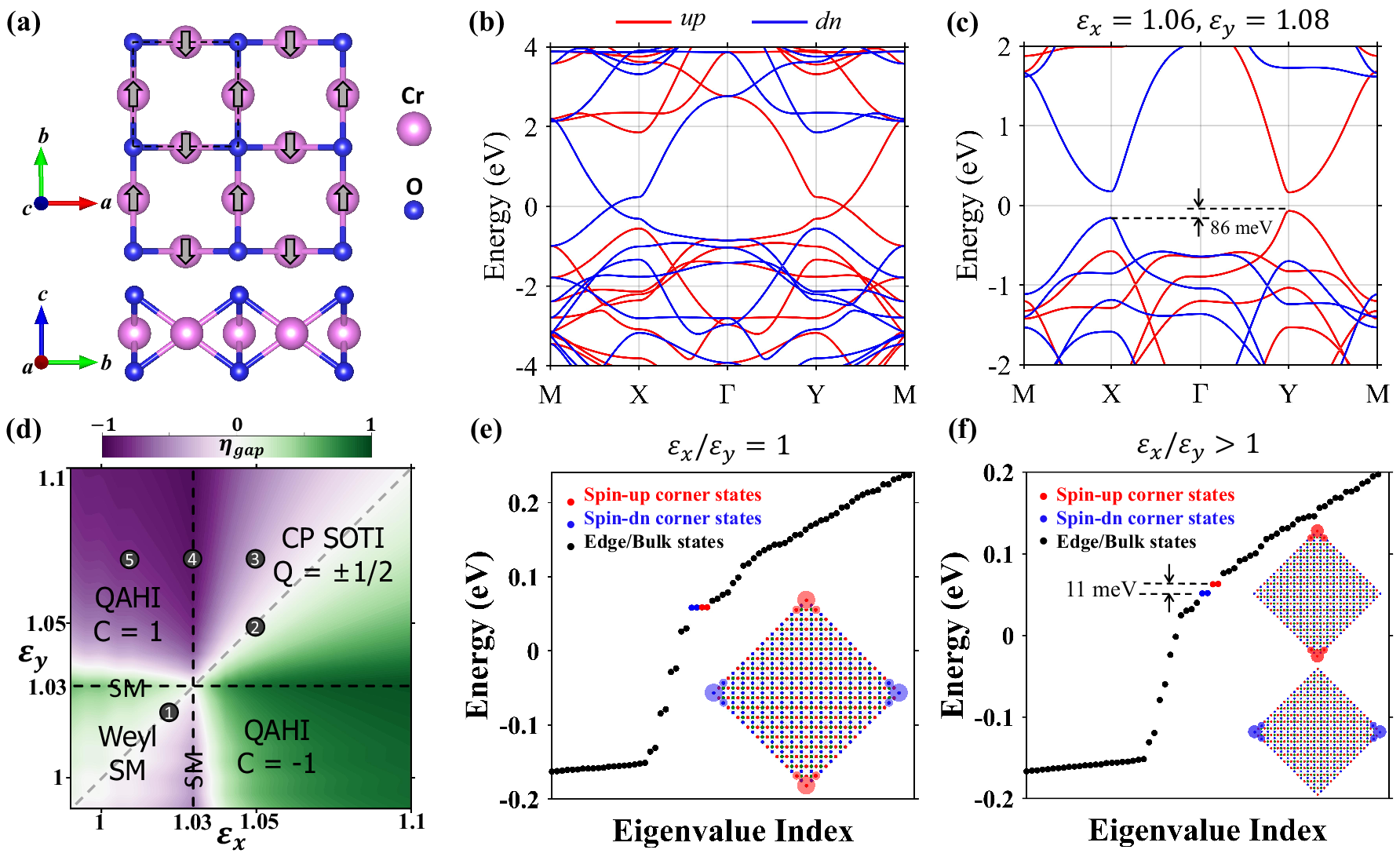}\caption{ 
		(a) Top and side views of CrO. Purple and blue spheres denote Cr and O atoms, respectively. (b) Spin-resolved band structure of CrO in the altermagnetic configuration. Red and blue represent spin-up and spin-down. (c) Band structure of CrO under anisotropic strain along the $x$- and $y$-axes. (d) Topological phase diagram under biaxial continuous strain, characterized by $\eta_{gap}$. (c) Based on the Wannier-based TB model, energy spectrum of the square quantum dot with preserved $M_{xy}$ symmetry, and (d) energy spectrum with broken $M_{xy}$ symmetry. The spin-associated corner states are marked in red and blue. Insets show the total charge distribution of the spin-resolved corner states.}\label{pho:m1}
\end{figure*}

After demonstrating the realization of spin-corner locking in the SOTI in 2D AMs, we now turn our attention to its realization in practical materials. First, we consider CrO as a candidate material for this property. It possesses a D$_{4h}$ point group symmetry, consisting of magnetic Cr atoms and non-magnetic O atoms. The magnetic Cr atoms adopt an antiferromagnetic configuration, as shown in the top and side views in Fig.~\ref{pho:m1}. The optimized lattice constant of monolayer CrO is $a = 3.23$ Å, with a Cr-O bond length of 2 Å. The collinear antiferromagnetic configuration, induced by the Néel vector arrangement, leads to a staggered magnetic antiferromagnetic phase. The intrinsic band structure, as depicted in Fig.~\ref{pho:m1} (b), shows staggered spin splitting along the $k_x$ and $k_y$ direction, with band crossing occurring at the X-M and Y-M points. At this stage, the system is in a Weyl semimetal phase. Upon breaking the space group symmetry through uniaxial strain, a QAHI phase with a nonzero Chern number is induced, this result previously discussed by Guo $et\ al.$~\cite{guo2023quantum}. By applying strain, the band merging can be eliminated, resulting in a gapped SOTI. It is noteworthy that when the strain is unequal along the two directions, mirror symmetry M$_{xy}$ or rotational symmetry $C_{4z}$ is broken, and the point group transforms to D$_{2h}$, leading to a fully compensated ferrimagnetism (FiM). This results in spin-polarized bands, as shown in Fig.~\ref{pho:m1} (c).
The Wannier-fitted Hamiltonian of CrO successfully reproduces the system’s band structure and spin characteristics, as shown in Fig. S2. Under the simplified Harrison rule, we obtain a linear first-order relation between the hopping parameters and the strain $\varepsilon$. Based on this, by tuning the hopping elements in the Wannier Hamiltonian, we construct the complete topological phase diagram of CrO as a function of $\varepsilon_x$ and $\varepsilon_y$, as shown in Fig. \ref{pho:m1} (d). The phase boundaries are distinguished by the spin-resolved band gap, defined as:
\begin{equation}
	\eta_{gap}=\frac{\Delta_{up}-\Delta_{dn}}{\Delta_{up}+\Delta_{dn}},
\end{equation}
where $\Delta_{up/dn}$ denote the band gaps for spin-up and spin-down channels, respectively. As shown in the phase diagram, when $\varepsilon_x = \varepsilon_y$ (along the gray dashed line), the system undergoes a transition from a Weyl semimetal to a spin-corner-locked SOTI. In contrast, when $\varepsilon_x \neq \varepsilon_y$, it leads to the emergence of CPSOTI and QAHI phases, corresponding to phases \textcircled{3} and \textcircled{5} in Fig. \ref{pho:m1} (d). The surface state of the QAHI is illustrated in Fig. S3.

Meanwhile, we also examine the symmetry protection of the second-order topological corner states. Under equal biaxial strain along the $x$ and $y$ directions, the system preserves both $M_{xy}$ and $C_{4z}$ symmetries, as well as the $C_{2z}$ symmetry. In 2D systems, higher-order topology is generally protected by crystalline symmetries such as rotation symmetry \cite{PhysRevB.99.245151}.
As shown in Figs. \ref{pho:m1} (e,f), the corner states persist even when the $C_{4z}$ symmetry is broken, indicating that the higher-order topological features are protected by the $C_{2z}$ symmetry. Accordingly, we evaluate the corner charge $Q_c$ under symmetry operations at high-symmetry points to characterize the topological nature of the system, as determined by the following expression \cite{PhysRevB.105.045417,PhysRevB.111.075433}:
\begin{equation}
	Q_{c}^{(2)}=\frac{e}{4}\left(-\left[X_{1}^{(2)}\right]-\left[Y_{1}^{(2)}\right]+\left[M_{1}^{(2)}\right]\right) \bmod e
\end{equation}
where $\left[\Pi_{\mathfrak{p}}^{(2)}\right]=\# \Pi_{\mathfrak{p}}^{(2)}-\# \Gamma_{\mathfrak{p}}^{(2)}$, the symbol "$\#$" denotes the count of symmetry eigenvalues at high-symmetry points. The eigenvalues of the $C_{2z}$ rotation symmetry are defined accordingly. Since $C_{2z}$ acts on the spin degrees of freedom, we calculate the spin-resolved corner charge $Q_c^{(2)} = e/2$ from all occupied states for spin-up and spin-down channels, respectively. These results confirm that monolayer CrO is a spin-resolved SOTI, carrying quantized fractional corner charges protected by $C_{2z}$ symmetry.
Building on previous studies of CrO \cite{guo2023quantum}, we explore the emergence of spin-selective second-order topological insulator (SOTI) phases and provide a detailed topological phase diagram that reveals the rich phase structure of the system.
Notably, this higher-order topological phase exhibits a strong and rare binding relation between spin and corner localization. By applying strain, we can energetically separate the spin-resolved corner states by up to 11 meV, which is sufficient for experimental identification and spin-resolved detection of cornertronic signatures. This mechanism differs fundamentally from the electric field tuning strategy used in Han $et \ al.$ \cite{PhysRevLett.133.176602}, where external fields were applied. In contrast, we utilize the system’s intrinsic symmetry and strain engineering to realize spin-resolved cornertronic functionality.

\begin{figure}[h]
	\centering
	\includegraphics[width=9.0cm]{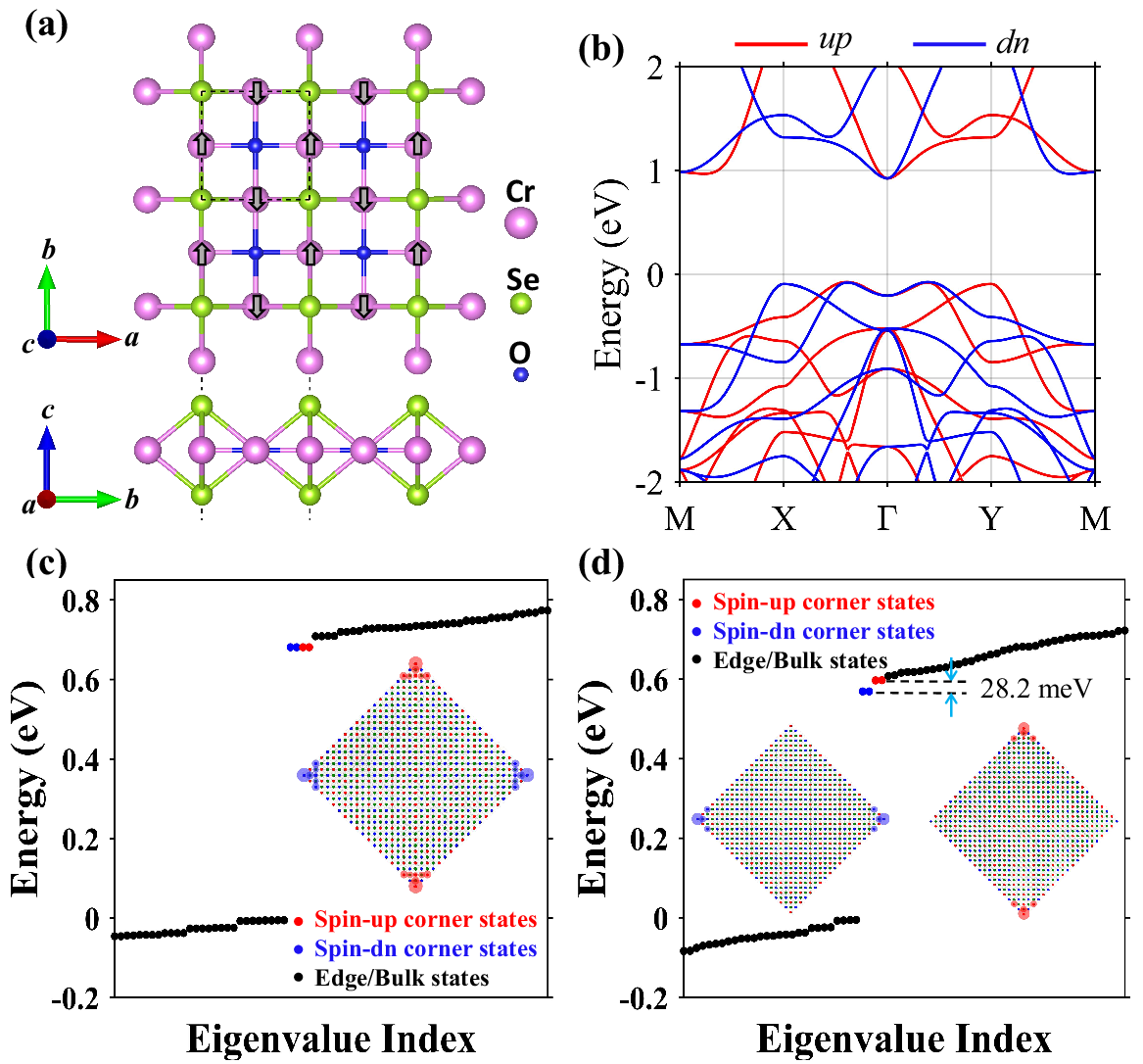}\caption{(a) Top and side views of Cr$_2$Se$_2$O. The purple, green, and blue spheres represent Cr, Se, and O atoms, respectively.
		(b) Band structure of the altermagnetic configuration of Cr$_2$Se$_2$O. 
		Based on the Wannier-based TB model, energy spectra of the corner quantum dot with preserved (c) and broken (d) $M_{xy}$ symmetry. Insets show the total charge distribution of the spin-resolved corner states.}\label{pho:m2}
\end{figure}

Based on the effective model, this type of 2D altermagnetic second-order topology exhibits a high degree of universality. Similar physics can be found in other altermagnetic materials that support corner-polarization second-order topology. For instance, as shown in Fig.~\ref{pho:m2} (a), Cr$_2$Se$_2$O serves as another candidate. Although it contains an additional nonmagnetic atom compared to CrO, they share the same antiferromagnetic configuration and space group symmetry. Likewise, the spin-resolved band structure of Cr$_2$Se$_2$O, displayed in Fig.~\ref{pho:m2} (b), reveals characteristic altermagnetic features.
By computing the parity eigenvalues at the high-symmetry points under the $C_{2z}$ rotation for all occupied states at the Fermi level, we again obtain a quantized fractional corner charge of $e/2$. in the absence of strain, the spin-polarized corner states are degenerate in energy, as shown in Fig.~\ref{pho:m2} (c). Upon applying a small uniaxial strain, these spin-resolved corner states become energetically distinguishable. Figure~\ref{pho:m2} (d) illustrates how distinct spin cornertronic features emerge at different energy levels.
It is worth noting that spin-polarized corner states have also been discussed in previous works. Gong $et\ al.$ realized such states by applying external electric fields \cite{gong2024hidden}. In contrast, our study induces spin polarization through the breaking of $M_{xy}$ symmetry and achieves spin-resolved corner electronics via a spin–corner locking mechanism intrinsic to 2D AMs. These approaches differ fundamentally in both their physical origin and implementation.

Here, we provide two representative material realizations. Recently, AB-penta systems have been reported to host CPSOTI phases \cite{wang2025pentagonal}; however, these systems do not exhibit spin-selective corner polarization. In contrast, the spin-selective CPSOTI phase proposed in our four-band model is is universal in nature, offering a platform for realizing spin-selective corner modes across diverse material systems.
The magnetic atoms may be substituted by other transition metals such as Mn or V, and the nonmagnetic atoms may include other chalcogens or halogens. Similar band structures indicating altermagnetic behavior have been reported in other materials, suggesting that they may also exhibit cornertronic phenomena as described in this work \cite{PhysRevB.111.134429,li2024strain}.
It is important to experimentally probe the second-order topological corner states in 2D AMs. The fabricated 2D AM quantum dots should preserve the $C_{2z}$ symmetry. The sample can then be placed on an Au(111) substrate. Following previous approaches for probing corner states \cite{hu2023identifying, kempkes2019robust}, scanning tunneling microscopy (STM) can be employed to perform current-imaging tunneling spectroscopy at the corner, edge, and bulk sites, measuring the differential conductance (dI/dV) at each location. The dI/dV spectra can then be used to distinguish the corner-localized states.

%\section{CONCLUSIONS}
In summary, we propose a class of altermagnetic SOTIs characterized by a strong coupling between spin and corner d.o.f.. This coupling leads to spin-resolved second-order corner states that exhibit distinct spin cornertronic features. Notably, the proposed spin-corner locked SOTI represents a rare and previously unexplored class among experimentally feasible SOTIs. The cornertronic d.o.f. is uniquely highlighted in the context of altermagnetism.
Moreover, based on first-principles calculations, we identify CrO and Cr$_2$Se$_2$O as material realizations of the proposed cornertronic SOTIs. This design strategy is not restricted to these two materials but is expected to be broadly applicable to a wide range of 2D altermagnetic systems with similar symmetry and magnetic configurations.
Our results not only uncover a new form of cornertronic behavior in SOTIs but also reveal a spin-corner-momentum entanglement in altermagnetic systems, thereby expanding the current understanding of magnetic topological phases.

%\subsection{References}

%\subsection{Methods}

%We explore the electronic characteristics of a hexagonal crystal-phase biphenylene lattice through the toolkit ATK’s first-principles calculations \cite{brandbyge2002density,ATK} and tight-binding models. In the first-principles calculations, we employ the Perdew–Burke–Ernzerhof (PBE) functional under the generalized gradient approximation (GGA), with a cutoff energy of 150 Ry and a 10×10×1 k-point grid. The lattice parameters and atomic positions are fully relaxed, with convergence criteria for energy and forces set at $10^{-5}$ eV and $10^{-3}$ eV/$\AA$, respectively. We employ a generously sized vacuum layer and incorporate at least 15 $\AA$ in the direction perpendicular to the electronic transport plane to prevent inter-layer interactions. In the calculation of light absorption, we use the TBPLaS software package \cite{li2023tbplas}, with a choice of 0.03 eV for the broadening of energy levels.

\begin{acknowledgement}
	This work is mainly supported by the National Natural Science Foundation of China (Nos. 11874113, 62474041) and the Natural Science Foundation of Fujian Province of China (No. 2020J02018).
\end{acknowledgement}

\begin{suppinfo}

Detailed DFT calculations, stability analysis, magnetic ground-state energies, spin-resolved second-order topological corner states, and further material characterizations

%The following files are available free of charge.
%\begin{itemize}
%  \item Filename: brief description
%  \item Filename: brief description
%\end{itemize}

\end{suppinfo}


%	\bibitem{ref01}   M. Z. Hasan and C. L. Kane, Topological insulators, Rev. Mod. Phys. {\bf82}, 3045 (2010).
%	\bibitem{ref02}   A. A. Burkov, Topological semimetals, Nat. Mater. {\bf15}, 1145 (2016).
%	\bibitem{ref03}   X. Ying and A. Kamenev, Symmetry Protected Topological Metals, Phys. Rev. Lett. {\bf121}, 086810 (2018).

%\bibliographystyle{plain}
\bibliography{References.bib}
%\begin{thebibliography}

\end{document}